# The Atrial Fibrillation Risk Score for Hyperthyroidism Patients

Ilya V. Derevitskii[1], Daria A. Savitskaya[2], Alina Y. Babenko[2], Sergey V. Kovalchuk[1]

[1] ITMO University, Saint Petersburg, Russia
[2] Almazov National Medical Research Centre, Institute of Endocrinology, St. Petersburg, Russia
ivderevitckii@itmo.ru

**Abstract.** Thyrotoxicosis (TT) is associated with an increase in both total and cardiovascular mortality. One of the main thyrotoxicosis risks is Atrial Fibrillation (AF). Right AF predicts help medical personal prescribe the correct medicaments and correct surgical or radioiodine therapy. The main goal of this study is creating a method for practical treatment and diagnostic AF. This study proposes a new method for assessing the risk of occurrence atrial fibrillation for patients with TT. This method considers both the features of the complication and the specifics of the chronic disease. A model is created based on case histories of patients with thyrotoxicosis. We used Machine Learning methods for creating several models. Each model has advantages and disadvantages depending on the diagnostic and medical purposes. The resulting models show high results in the different metrics of the prediction of AF. These models interpreted and simple for use. Therefore, models can be used as part of the support and decision-making system (DSS) by medical specialists in the treatment and diagnostic of AF.

## 1 Introduction

Thyrotoxicosis (TT) is associated with an increase in both total and cardiovascular mortality. The majority of patients with TT are working-age individuals. Consequently, the negative social impact of TT remains incredibly significant [1]. Due to the risk of thromboembolic events and heart failure, one of the severest complications of TT is atrial fibrillation (AF). The thyrotoxic AF (TAF) incidence is as follows: 7–8% among middle-aged patients, 10–20% in seniors and 20–35% for those having ischemic heart disease or valvular disease [2–5]. High incidence, severity and also a significant adverse impact on life quality – all these facts make TAF prevention a crucial problem. Regrettably, there is no established TAF risk calculation scale at the moment. Nevertheless, devising one would allow selecting patients with a high risk of TAF for the closest follow-up or for early radical treatment of TT – surgical or radioiodine therapy instead of long-term medical treatment. This will ultimately lead to a decrease in TAF frequency.

## 2 Existing methods overview

In the literature, the topic of assessing the risk of AF development is described in sufficient detail. There are several approaches to solving this problem. In the first approach, the authors study risk factors for atrial fibrillation, predicting the probability of AF development for a specific period using regression analysis. For example, the 10-year risk of AF development is estimated[1]; multivariate Cox regression is used as tools. The results obtained with this approach are quite reliable, but such models do not include many factors that presumably affect the risk of AF. Also, these models do not take into account the specifics of AF development in patients with certain chronic diseases (in contrast to the models of the second approach). In the second approach, the risk of AF is estimated for patients with a certain chronic disease, such as diabetes[5], and it is revealed that patients with diabetes have a 49% higher risk of developing AF. In the third approach, meta-studies of populations are conducted to analyze the risk factors for AF, for example[6]. In this article, the authors described a cross-sectional study of California residents. Those results are interesting and can be used to calibrate the models obtained by MO methods; however, within using these works we cannot assess the probability of AF development in a particular person. Many factors can influence the probability of AF. It is necessary to consider both the specifics of this complication and the specifics of a chronic disease when we create a model for assessing the risk of AF. Therefore, special methods are needed for creating such a model. This study presents an approach to the estimator's risk of atrial fibrillation for patients who suffer from thyrotoxicosis. This approach considers the specifics of the influence of the thyrotoxicosis course for a particular patient on the probability of AF.

To date, a lot of data have been obtained about thyrotoxic atrial fibrillation risk factors. The majority of investigations have demonstrated that the most significant predictors of this severe arrhythmia are advancing age, male gender, prolonged thyrotoxicosis duration, high level of thyroid hormones and concomitant cardiovascular diseases[2, 7–11]. Findings from other researches have shown that factors like female gender, obesity, heart rate more than 80 beats per minute, the presence of left ventricular hypertrophy, big left atrium diameter, chronic renal disease and proteinuria, elevated liver transaminase and C-reactive protein levels predispose to thyrotoxic atrial fibrillation[12–14]. Thus, based on the comprehensive literature review, we can conclude that available data on thyrotoxic atrial fibrillation predictors are contradictory and, furthermore, a lot of studies have an insufficient evidence base. Moreover, for an individual's absolute risk evaluation, integrating multiple risk factors and identify the most significant of them, are required. Accordingly, it is necessary to develop a system for ranking risk factors by their degree of impact and models for calculating the risk of thyrotoxic AF. Using the system for ranking we need to create an easy-to-use and interpretable model for assessing the risk of developing atrial fibrillation in patients with thyrotoxicosis. All of the above explain relevance of this study.

# 3 Case study

## 3.1 Data

The study included 420 patients with a history of overt thyrotoxicosis. The study sample was divided into two subgroups: one includes 127 (30.2%) individuals with thyrotoxic atrial fibrillation and other - 293 (69.8%) patients without. All participants were treated previously or at the time of the study for thyrotoxicosis in Almazov National Medical Research Centre or in Pavlov First Saint Petersburg State Medical University, St. Petersburg, Russia in 2000-2019.

Entry criteria:

- Men and women with overt thyrotoxicosis, associated with Graves' disease, toxic adenoma or multinodular toxic goiter.
- Age between 18 and 80 years.

Exclusionary criteria:

- Subclinical hyperthyroidism.
- A history of atrial fibrillation developed before the onset of thyrotoxicosis.
- Concomitant diseases that may affect structure and function of the cardiovascular system: hemodynamically significant valve disease, nonthyrotoxic cardiomyopathy, severe obstructive lung diseases.
- Chronic intoxication (alcohol, toxicomania).
- Pregnancy at the time of thyrotoxicosis.

The data includes 4 groups of features (abbreviations are indicated in brackets):

- Physiological: gender, age, height, weight, body mass index (BMI), the genesis of thyrotoxicosis «Genesis of TT», smoking status «SC», total cholesterol «TC», triglycerides «TG», low density lipoprotein cholesterol «LDL» level high density lipoprotein cholesterol «HDL», creatinine «CR», potassium «K» and hemoglobin «HB» levels during TT.
- Initial cardiological status before the development of TT: arterial hypertension «AH before TT», coronary heart disease before TT «CHD before TT», various heart rhythm disorders before TT «HRD before TT», congestive heart failure «CHF before TT», heart rate reducing therapy antiarrhythmic drugs, ivabradine, beta-blockers, and other types of therapy «HRRT before TT».
- Characteristics of thyrotoxicosis: free tetraiodothyronine «FT4», triiodothyronine «FT3», thyroid-stimulating hormone receptor antibodies level «TSHRA», duration of TT, for patients with AF - duration of TT before AF «DTT», subclinical TT duration (more less than 12 months) «DSTT», the number of relapses of TT «RTT», the presence of episodes of hypothyroidism «EHT», therapy of thyrotoxicosis «TTT», the duration of thyrostatic therapy «DTST», the time after TT onset when thyroidectomy or radioiodine therapy was provided «TTRTP».
- Cardiological status during the TT: arterial hypertension «AH during the TT», heart rate during the first examination «HR during the TT» and maximum recorded during TT heart rate «MHR during the TT», supraventricular extrasystole «SVE during the TT», ventricular extrasystole «VE during the TT», other heart rhythm disorders «OHRD heart rhythm disorders», congestive heart failure «CHF during the TT», heart rate reducing therapy «HRRT during the TT».

### 3.2 Models creation

The main objective of the study is to create a practical method for assessing the risk of developing atrial fibrillation in a patient with thyrotoxicosis. The model should include both features unique to the disease and physiological parameters.

At the initial stage, we have a table with observations(rows) and features(columns). The preprocessing of this dataset included five-stage. This stage shows in figure 1.

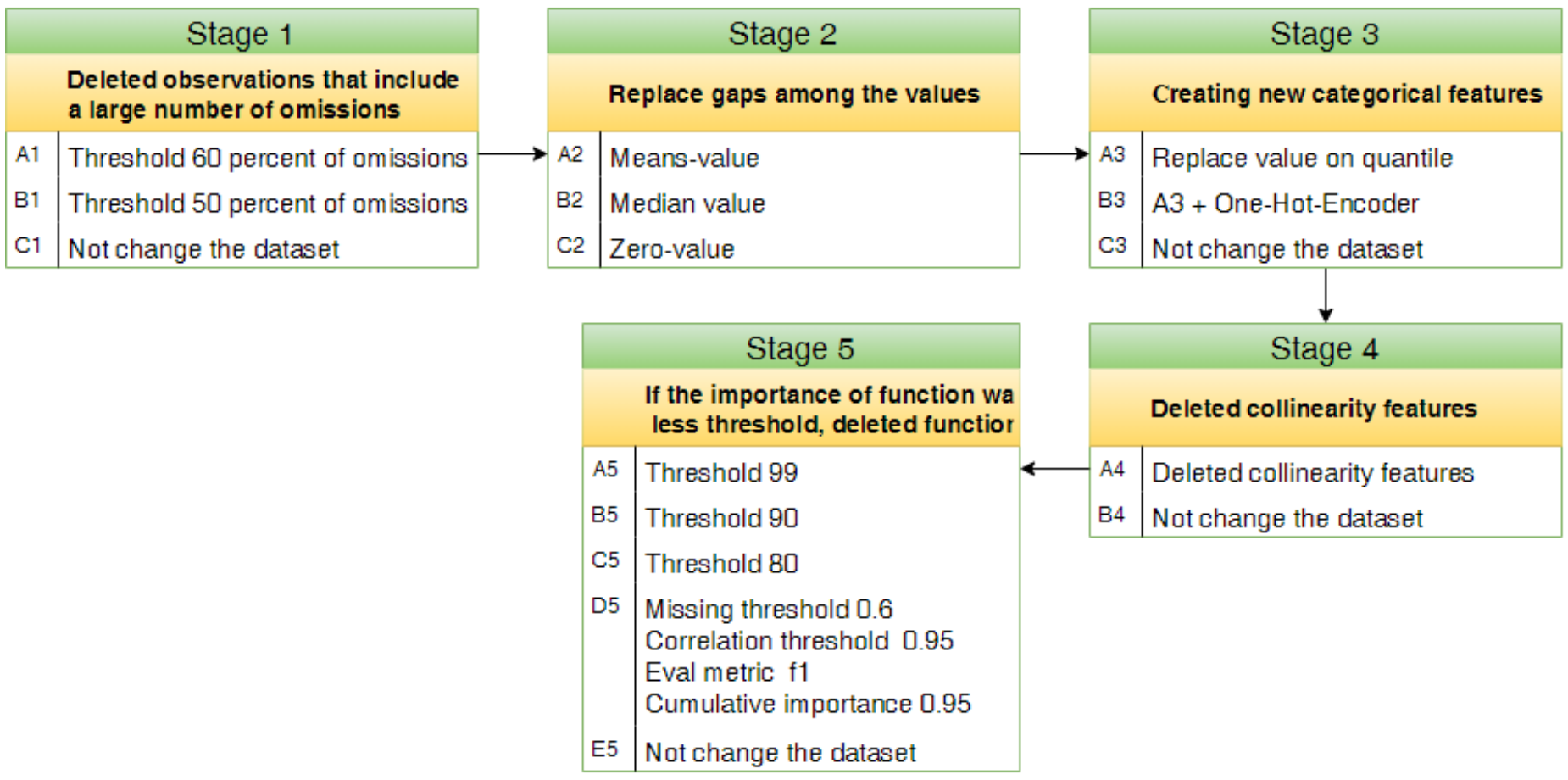

**Fig. 1.** Preprocessing stages

In the first stage, we deleted observations that include a large number of omissions. In the second stage of preprocessing we replace gaps among the values. In the third stage, we calculated quantiles for each discrete feature. Next, we divided the values of all discrete features into three intervals. We use quantiles as intervals threshold. Using the B3 method, we first use the A1 method. Then we use One-Hot-Encoder methods for coding categorical values. On the fourth stage, we tested each featured couple on collinearity. On the next stage, we selected features for each dataset.

Finally, we created 180 different datasets (2*2*3*2*5). Each dataset contains a unique sequence of processing and selecting variables for prediction.

Next, the sample was divided into test and training, classes were balanced by pre-sampling features. The model was selected from the models KNeighborsClassifier, RandomForestClassifier, LogisticRegression, DecisionTreeClassifier, XGBClassifier, SVC, GaussianNB, MultinomialNB, BernoulliNB. For each model, metrics f1, accuracy, recall, precision for cross-validation were calculated for each of the datasets. Table 1 is a fragment of a table with the results of cross-validation.

**Table 1.** Models comparison

| Model | Accuracy averege | F1 averege | Recall averege | Precision averege | Preprocessing |
|---|---|---|---|---|---|
| XGB | **0.876** | **0.793** | 0.788 | **0.801** | B1 A2 C3 B4 B5 |
| BernoulliNB | 0.834 | 0.742 | 0.788 | 0.701 | A1 C2 B3 B4 A5 |
| GaussianNB | 0.819 | 0.73 | **0.8** | 0.677 | B1 C2 B3 B4 C5 |
| LR | 0.835 | 0.726 | 0.715 | 0.742 | A1 C2 B3 B4 B5 |
| DecisionTree | 0.806 | 0.703 | 0.675 | 0.666 | B1 A2 A3 B4 B5 |
| RandomForest | 0.819 | 0.687 | 0.609 | 0.753 | C1 B2 B3 A4 C5 |
| MultinomialNB | 0.777 | 0.653 | 0.693 | 0.618 | A1 C2 B3 B4 C5 |
| KNeighbors | 0.796 | 0.614 | 0.537 | 0.720 | A1 C2 B3 A4 B5 |
| SVC | 0.765 | 0.452 | 0.327 | 0.778 | A1 C2 B3 B4 D5 |

This table includes the best combination of dataset preprocessing steps and model type for each model. The table is ranked by the f1 average. The best result was shown by the XGBClassifier model - 0.79 f1 average, 0.88 accuracies, and 0.8 precision. The method of processing the dataset is to remove features with more than 50% gaps, close gaps with averages, do not encode variables with naive coding and do not use quantiles, do not delete collinear features, the threshold for selecting functions is 0.9. Given the initial balance of classes 3 to 7 with a given quality, it makes sense to apply this model in practice. In terms of recall, the best result was shown by the GaussianNB model - 0.8. For reasons of interpretability, it is also advisable to use the DecisionTreeClassifier model (0.81 accuracies and 0.70 f1). Also, a good result was shown by the LR model (0.835 accuracies). This model has high interpretability. Also, for the convenience of using this model, it can be translated into a simple scale. To each attribute to associate a certain score. Given the high sensitivity and specificity, it also makes sense to use it in practice.

The main study objective is the creation of a set of models convenient for their application in practice. Today, there are many predictive models, but fewer of them are applied in practice. This is due to many factors. For example, the reluctance of doctors to work with the new software on which the model is based. Or the model has high accuracy, but it is not clear how to interpret the results. The decision-making system in the form of a scale is understandable and convenient for medical specialists. Each factor has a specific score. This score indicates the contribution of the factor to prognosis. Scales are widespread in medical practice, for example, the FINDRISK scale [15].

On base Logistic Regression model, we can create a scale for estimated atrial fibrillation risk. The preprocessing process is shown in figure 1. Out of 176 signs, the model showed the best quality at 60. The grid search method found model parameters.

The features were awarded points based on the regression coefficient. If the absolute value of the regression coefficient lay in the half-interval (0.0.5] - 0.5 points, (0.5.1] -1 point, (1.2] - 2 points, (2.3] - 3 points, (3.4] - 4 points.

The following values most significantly reduce the risk of atrial fibrillation: supraventricular extrasystole, which appeared against the background of thyrotoxicosis, duration of thyrotoxicosis is less than 6 months (coefficient -4); atrial pacemaker migration, Infiltrative ophthalmopathy and pretibial myxedema, duration of thyrotoxicosis of more than 6 and less than 12 months (coefficient -3); lack of arterial hypertension in the presence of thyrotoxicosis, presence of hypertension with target blood pressure values during treatment, hemoglobin more than 120 g / l and less than 132 (coefficient -2).

The most significant increase in the risk of Atrial Fibrillation is as follows: Diabetes mellitus (+4 coefficient); Heart rate on the background of TT is greater than 89 and less than 95 (coefficient 3); thyroidectomy thyrostatic therapy of thyrotoxicosis, male, lack of pulse-reducing therapy against thyrotoxicosis, impaired glucose tolerance, ventricular extrasystole, which appeared against the background of thyrotoxicosis, duration of tirputeostatic therapy, months> 40.0, time of radical treatment (how many years after the debut thyrotoxicosis performed thyroidectomy or radioiodine therapy, months) less than 77(+2 coefficient).

The developed scale was designed in the form of a questionnaire, which allows, without medical tests, to assess the risk of developing Atrial Fibrillation for a particular patient. You can see a fragment of this questionnaire in Figure 2.

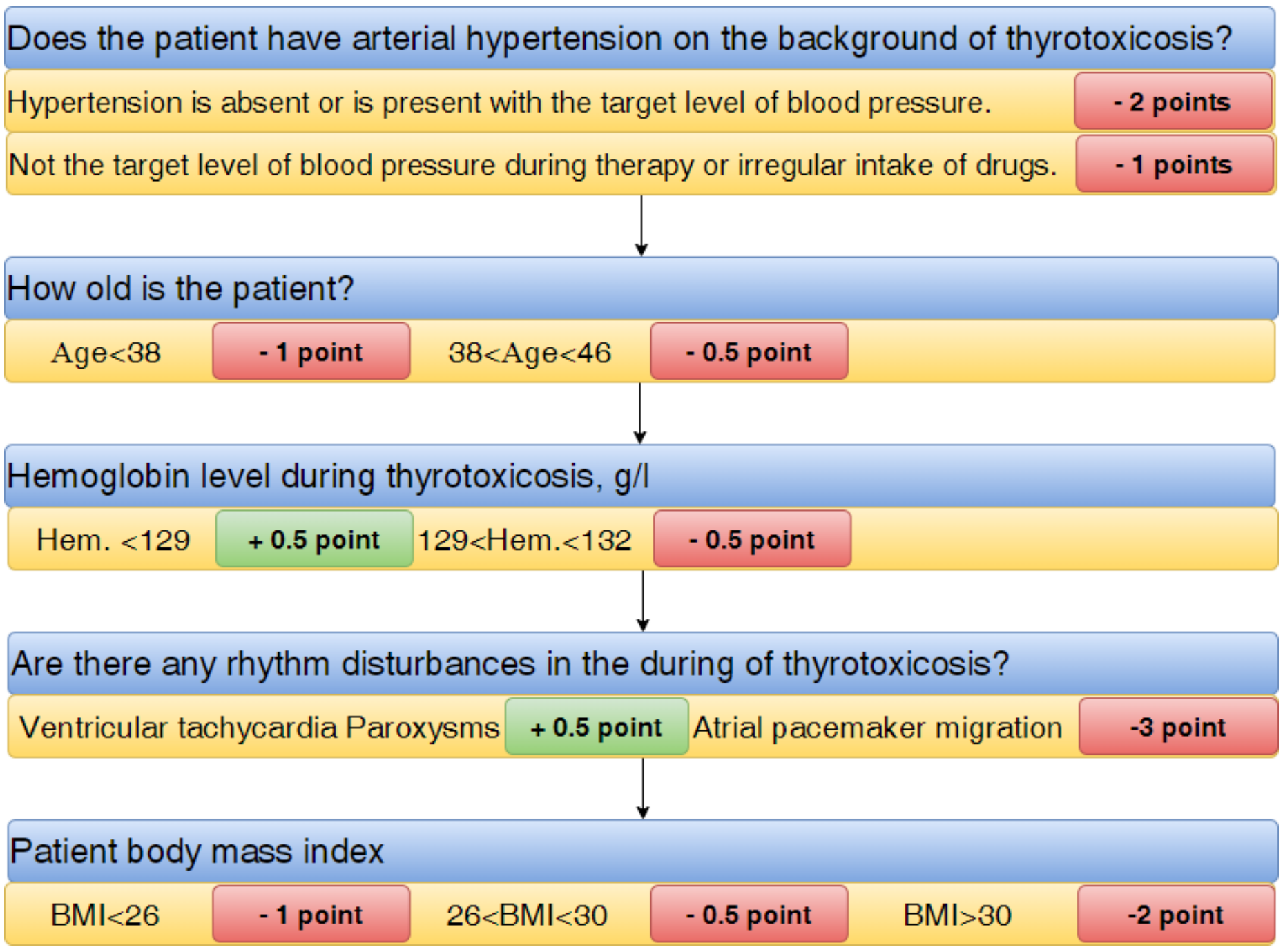

**Fig. 2.** Fragment of AF risk questionnaire

The final score was calculated for each patient, patients were divided into 4 risk groups. Table 2 shows the incidence of AF. Groups 2 and 3 include the most severe patients, the indicator of getting into this group shows very high values of sensitivity and specificity. This table shows the accuracy of the created scale. This scale is a standard scale for assessing the risk of complication, created using the same tools as the well-known FINDRISK scale [15] and was validated similarly[16, 17].

**Table 2.** Risk scale metrics for the development of AF.

| Risk level | frequency_FP0 | frequency_FP1 | Score |
|---|---|---|---|
| Low | 1 | 0 | <=-5 |
| Average | 0.454545 | 0.545455 | (-5,1] |
| Tall | 0.0980392 | 0.901961 | (1,5.5] |
| Very tall | 0.0526316 | 0.947368 | >=5.5 |

### 3.3 Results analysis

In figures 3 and 4, we visualized the 1st tree in gradient boosting.

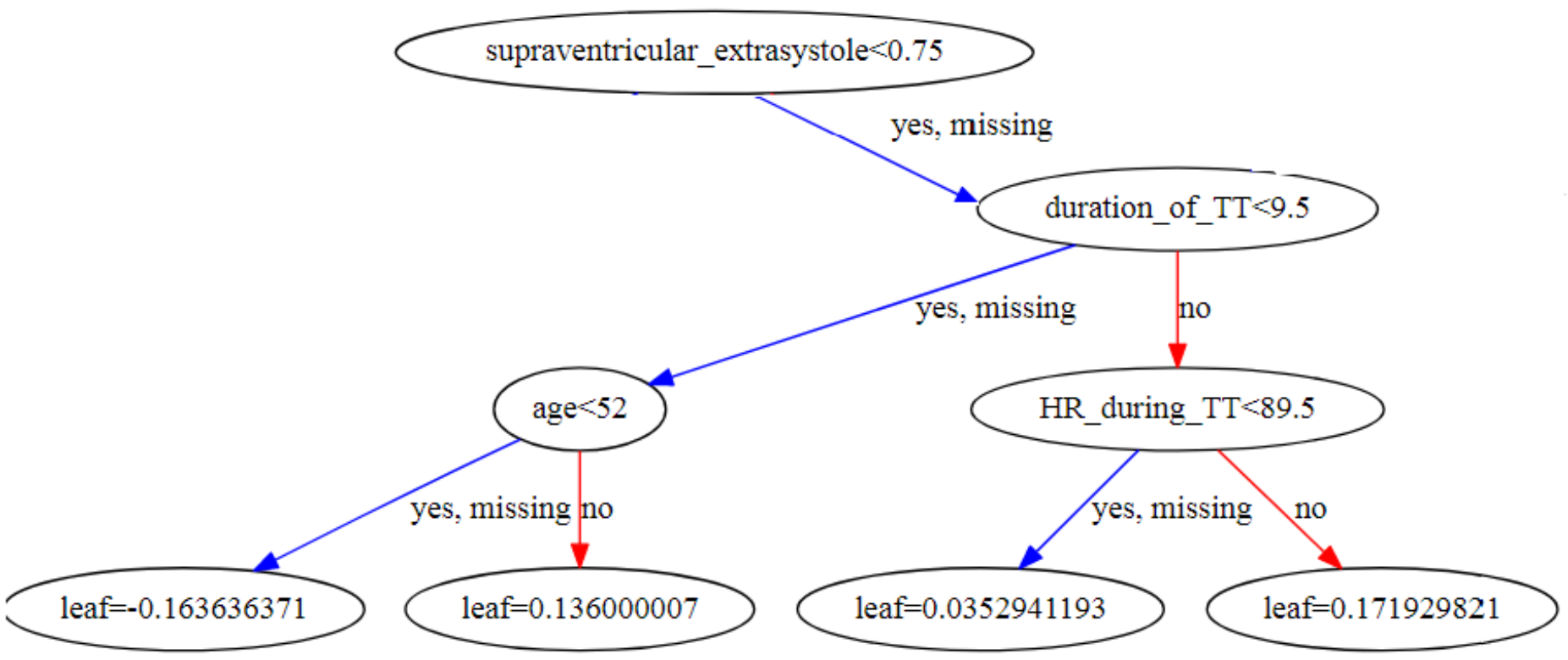

**Fig. 3.** Fragment of decisions tree for patients with supraventricular extrasystole

Figure 3 we visualized a fragment of decision tree for patients with supraventricular extrasystole. Presumably, the most important symptom is the presence of supraventricular extrasystole in the patient. Also, this decision tree confirms the a priori significance of the signs describing the specificity of the course of thyroid disease in a particular patient. Presumably, the longer the duration of thyrotoxicosis, the higher the probability of AF. The age and heart rate detected at the first examination also increase the probability of atrial fibrillation in patients with thyrotoxicosis and supraventricular extrasystole.

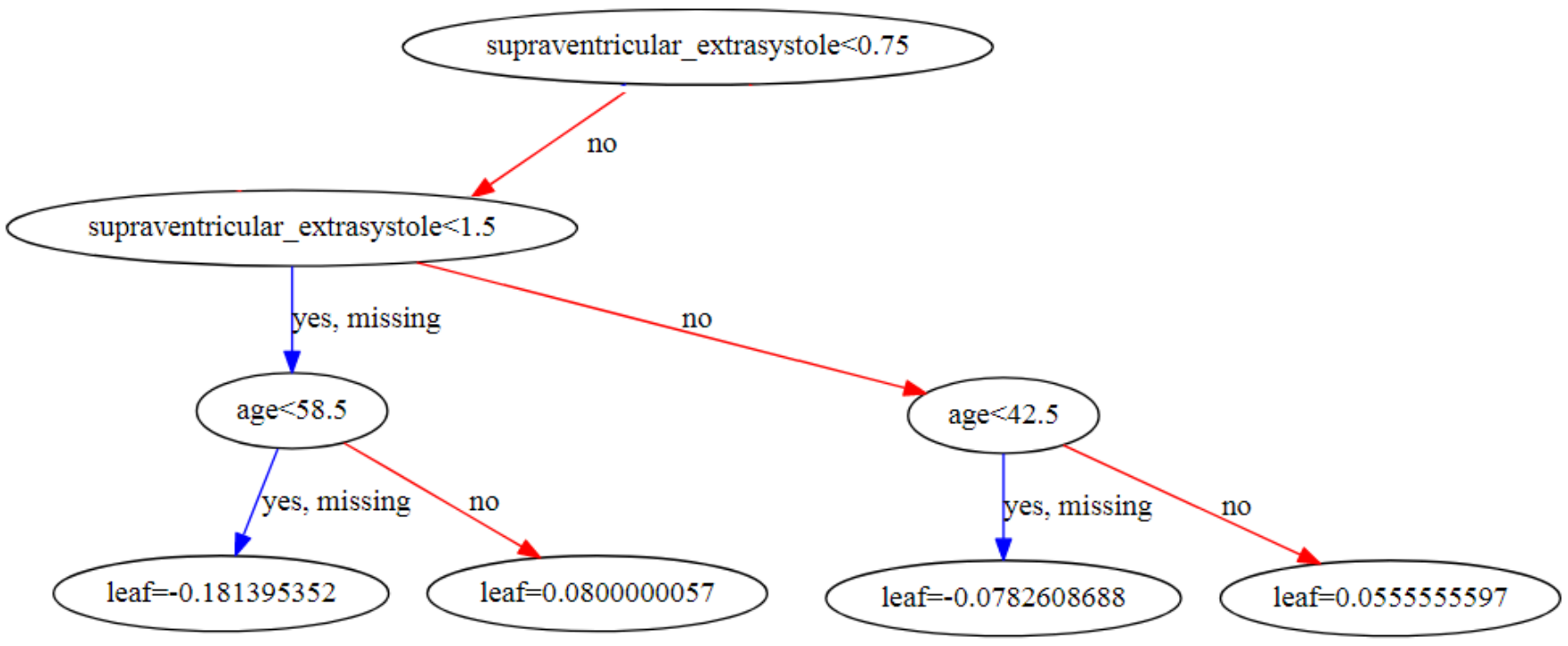

**Fig. 4.** Fragment of decisions tree for patients with supraventricular extrasystole

Figure 4 we visualized a fragment of decision tree for patients without supraventricular extrasystole. From the first tree, it follows that the most important criterion for the presence of AF is age. However, the specifics of constructing trees in gradient boosting does not always allow us to determine the importance of a particular trait for a predicted variable. We try to analyze the decision tree created on the basis of the DesissionTreeClassifier model. Visualization of 2 fragments of this tree is presented in figures 5 and 6.

Figure 5 describes the branch for patients who do not have supraventricular extrasystole. Based on the tree structure, supraventricular extrasystole is one of the most important signs that contribute to the probability of AF. The absence of supraventricular extrasystole with an 89% probability means that AF will not develop in this patient. Further, patients are divided by age, threshold 58.5. Age above 58 indicates a high probability of developing AF if in these patients the FDT3 level exceeds 1.6. For patients younger than 58 without extrasystole, the probability of developing AF is high only if thyrostatic therapy duration is above 225 or the duration of thyrotoxicosis is above 65 months. The risk of developing thyrotoxicosis is also affected by body mass index (the higher the variable, the higher the risk of AF) and pulse therapy. If therapy was carried out, the risk of atrial fibrillation is significantly higher than in the absence of therapy.

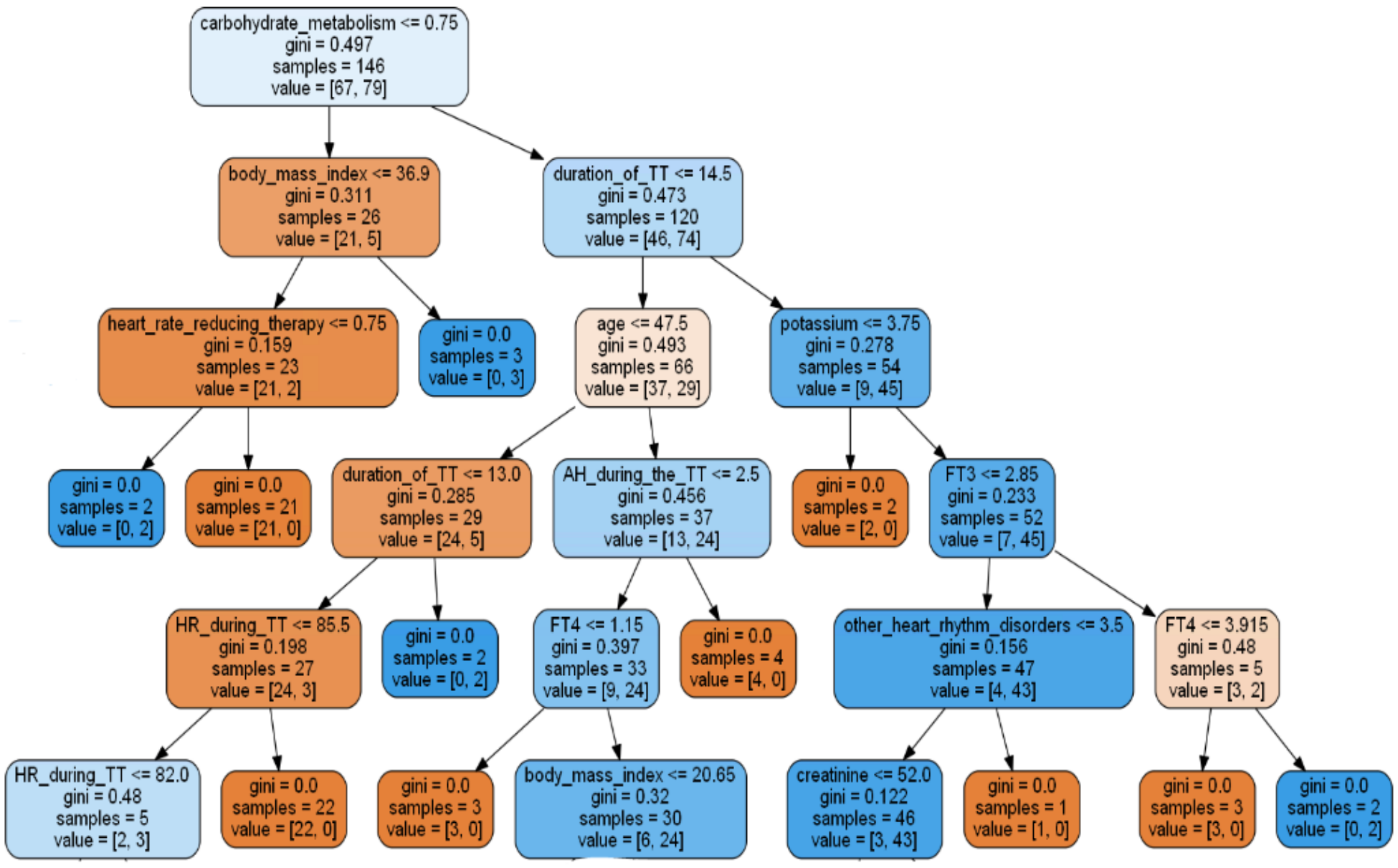

**Fig. 5.** Fragment of decisions tree for patients without supraventricular extrasystole

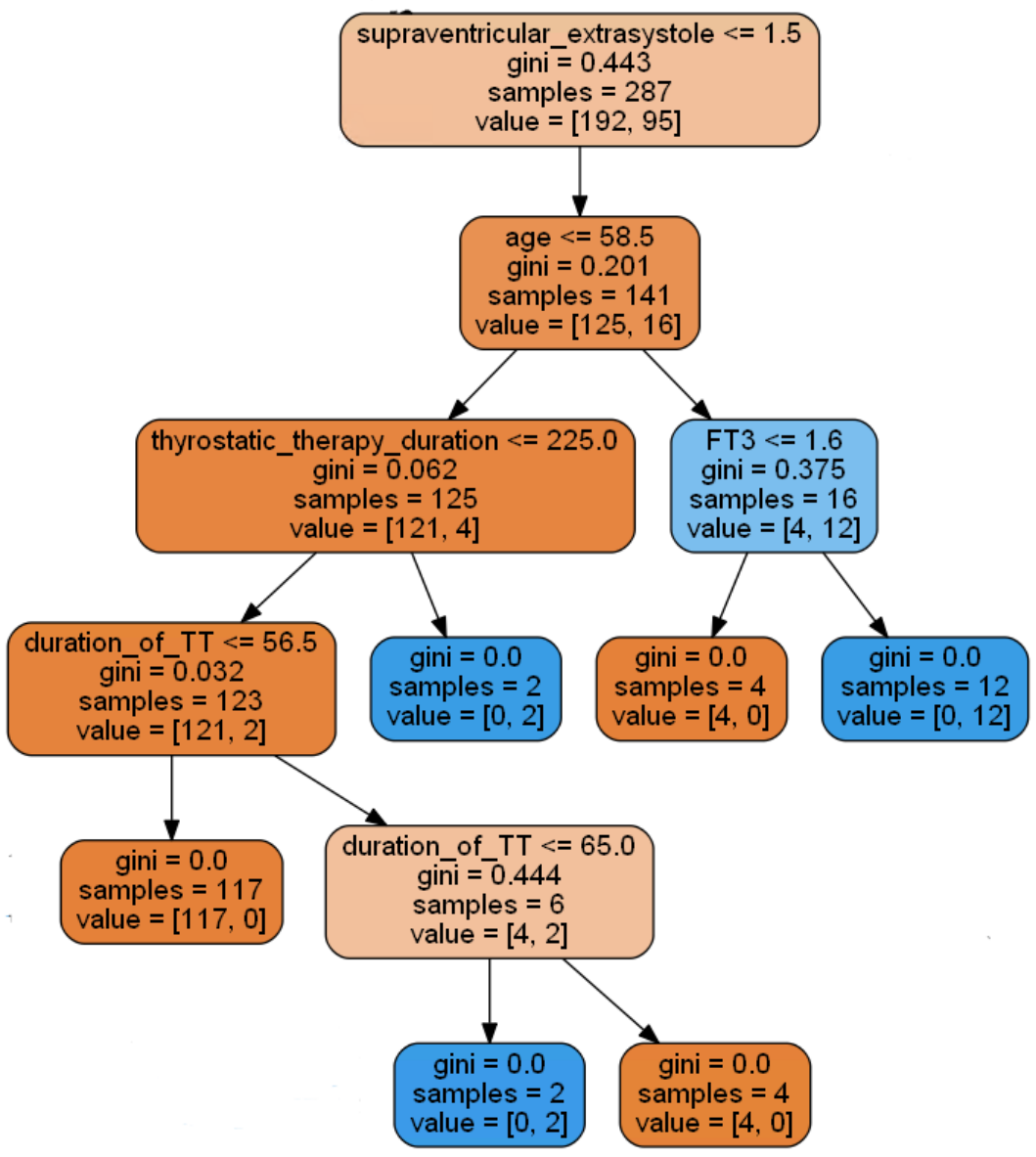

**Fig. 6.** Fragment of decisions tree for patients with supraventricular extrasystole

We analyzed the second fragment - the course of the disease for patients with supraventricular extrasystole. For this group of patients, obesity of the 2nd and 3rd degrees indicates a high probability of A even in the absence of problems with carbohydrate metabolism. If the patient does not have obesity, does not have a violation of carbohydrate metabolism, and does not undergo heart rate reducing therapy, the probability of AF is close to 0. If there are problems with MA without analyzing other factors, the patient has a 60% chance of developing AF. Further risk assessment depends on age (the younger the patient, the lower the risk), potassium level, FT3, the presence of other heart rhythm disturbances, FT3 and the duration of hypertension. These findings are confirmed by a priori knowledge and expert opinion of doctors. This decision tree is a fairly simple and well-interpreted method that can be useful as part of a support and decision-making system for medical professionals working with patients with thyrotoxicosis. Similarly, the complete structure of the tree was analyzed. The findings were interpreted by experts in the field of thyrotoxicosis. The knowledge gained about the relationship between the risk of AF and the specific course of thyrotoxicosis can be used to create a support and decision-making system.

Confusion matrices were constructed and analyzed. In terms of f1, XGBoostClassifier has the best predictive power. However, Logistic Regression has fewer false-negative errors. Given the balance of classes, DecisionTreeClassifier has good predictive power, despite the simplicity of the model (sensitivity 0.8 and specificity 0.63). DecisionTreeClassifier is the most convenient model for interpretation. The choice of a specific model depends on the goals that the medical specialist wants to achieve. If it is more important to find all patients at risk of developing AF, XGBoost is better, because the model has the most specificity. If it is important to find all patients whose AF does not develop, then it is more logical to apply Logistic regression. For ease of use and interpretation of the result, it makes sense to use DecisionTreeClassifier or scale.

## 4 Analysis atrial fibrillation paths in dynamic

We analyzed at the way atrial fibrillation occurs. We used data from 1450 patients suffering from thyrotoxicosis of any form. We divided the patient data into 2 groups. Group 1 includes 450 patients who have developed atrial fibrillation. Group 2 includes 1000 patients in whom atrial fibrillation did not develop in the observed period. Next, we present the patient's course of diseases in the form of a pathway. We use graphs representation. The nodes were the codes of "mkb10" (international statistical classification of diseases) with that the patient was admitted for treatment to the Almazov Medical Center. We created edges from an earlier node to a later node. In figures 7 and 8 showed graphs for the two groups. The colors show the clusters. These clusters were obtained by the method of maximizing modularity. The size of the node depends on its degree.

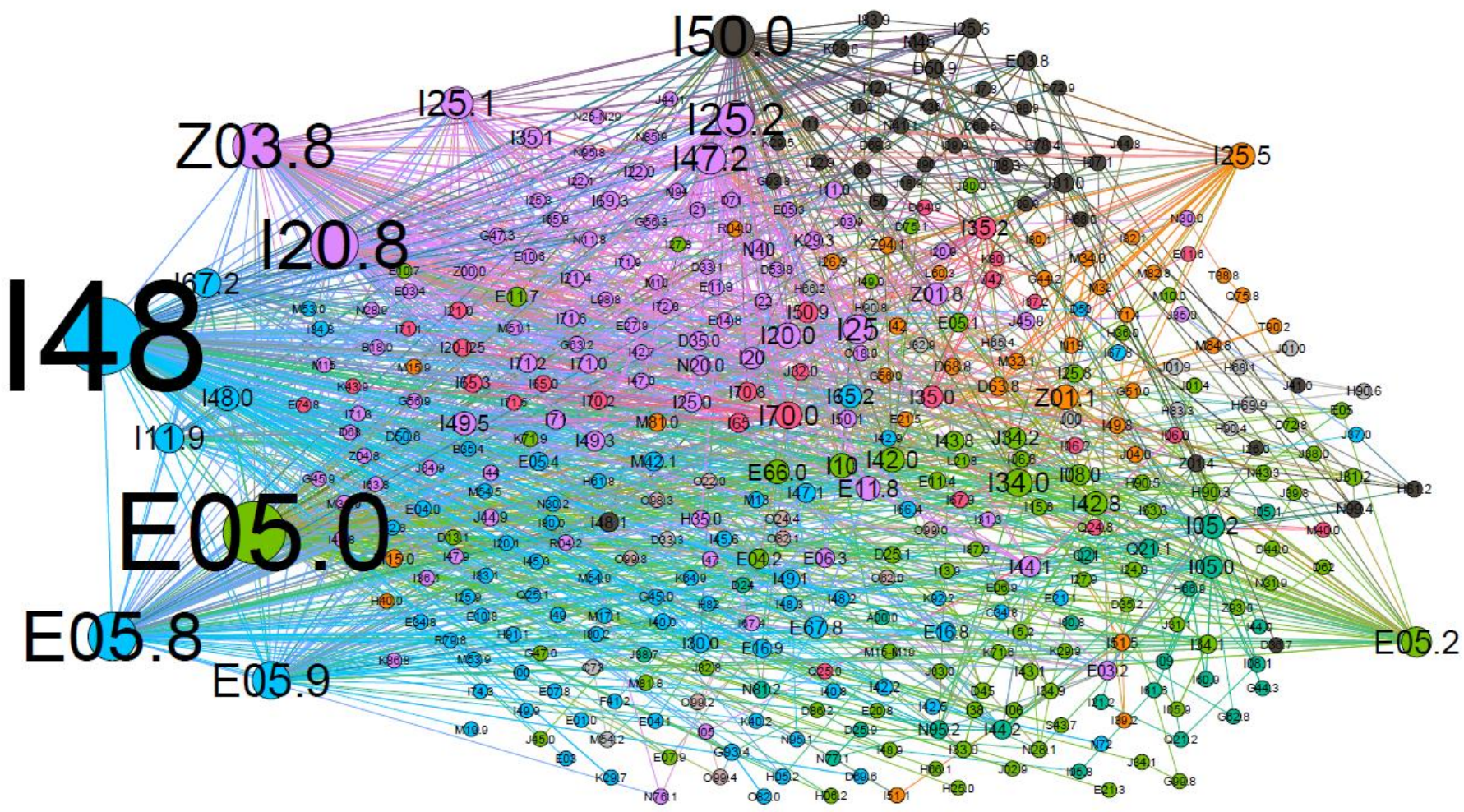

**Fig. 7.** Pathways of thyrotoxicosis for 450 patients with atrial fibrillation

This graph contains 211 nodes "mkb10" and 1035 edges. Largest nodes in terms of degree are observation for suspected other diseases or conditions (Z03.8, purple cluster), congestive heart failure (I50.0, black cluster), past myocardial infarction ( I25.2 purple cluster), other forms of angina pectoris (I20. 8, purple cluster), hypertension (I10, green cluster), atherosclerotic heart disease (I25.1, purple cluster), past myocardial infarction (I25.1, orange cluster). A diagnosis of AF is associated with each of these conditions. Therefore, when one of these conditions occurs, the patient is at risk of atrial fibrillation. It is a signal for the use of the above models. We include all large nodes in the model features.

Heart failure diagnosis is the center of gray clusters. This disease relates TT with another diagnosis in gray clusters. This cluster includes several types of anemia, heart disease, and limb disease. We assume that these diseases influence on risk AF only together with heart failure diagnosis. Observation with suspected other diseases or conditions, other forms of angina pectoris, acute transmural infarction of the lower myocardial wall, ventricular tachycardia are the largest nodes in the purple cluster. This cluster in heart disease clusters. All these diseases were included in model train features. The pathways of AF for patients of 2 types of thyrotoxicosis are described in the green cluster. These species are thyrotoxicosis with diffuse goiter and thyrotoxicosis with toxic multinodular goiter. If patient suffers from these diseases, his risk-AF-indicators include codes E66.0 (obviosity) I10 (diabetes mellitus), cardiomyopathies (I42.0,I42.8). Therefore, for these patients group right diets is very important. Blue cluster included patients which other forms of thyrotoxicosis and thyrotoxicosis unspecified. For patients in this pathway group, indicators of AF include hypertension, cerebral atherosclerosis. Since 2 types of atrial fibrillation belong to this cluster, patients from this group must be automatically spent in the risk group.

The following types of atrial fibrillation fall into the blue cluster: Paroxysmal atrial fibrillation, typical atrial flutter. These forms of AF are most closely associated with a diagnosis of E05.8 (unspecified forms of thyrotoxicosis), E05.9 (unspecified forms of thyrotoxicosis). All clusters were analyzed. And we identified indicators of increased risk for each particular type of Atrial Fibrillation. Those pieces of knowledge used in predictive models of atrial fibrillation.

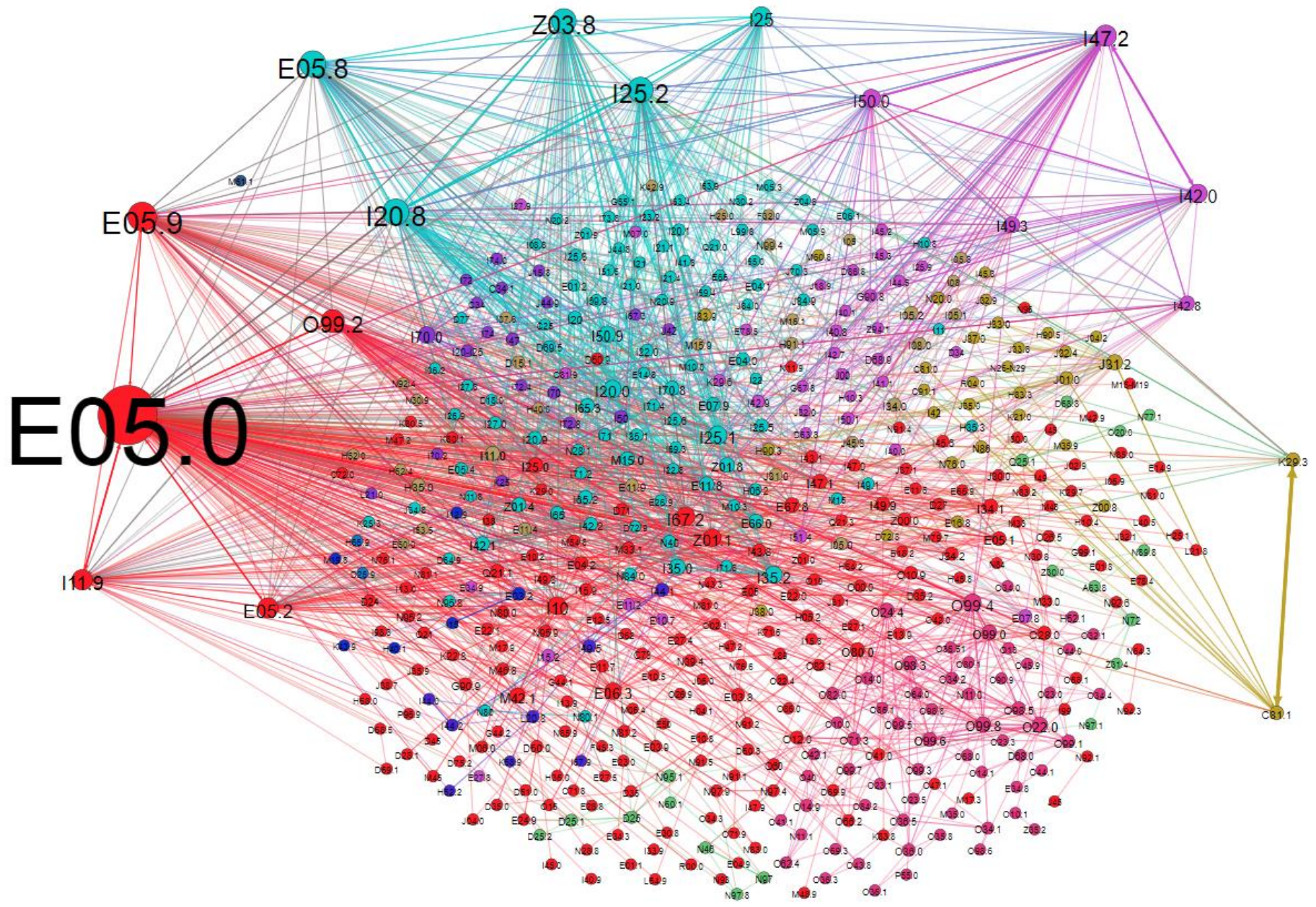

**Fig. 8.** Pathways of thyrotoxicosis for 1000 patients without atrial fibrillation

**Table 3.** Graph Statistics.

| Statistic | Graph for first group | Graph for second group |
|---|---|---|
| Average degree | 4,52 | 5.355 |
| Graph diameter | 10 | 9 |
| Graph density | 0.009 | 0.014 |
| Components | 9 | 4 |
| The average length of the path | 3.217 | 3.09 |
| Average clustering factor | 0.308 | 0.33 |

Based on table 3, the graph structure is different. The average degree is higher in the trajectories of patients without AF. This can be explained by different group sizes. The different structure of the course of thyrotoxicosis for these groups of patients follows from a different number of components. We can calculate the distance of similarity for a couple of particular pathways. Therefore, we propose new features - distance from particular pathways to pathways from these graphs. We think, when this feature is included in the Machine Learning model, the quality of the predicted is improved. We plan to study this hypothesis in future work. The obtained knowledge will be used in future work to create a dynamic model that predicts the risks of other complications.

## 5 Conclusion and Future Work

As a result of the study, we created a practical method for assessing the risk of AF. This method includes 4 models. The choice of a particular model by a medical specialist depends on the specific treatments-diagnostic goal. The best combination of specificity and completeness was shown by a model based on Gradient Boosting. This model takes into account both the specifics of the course of thyrotoxicosis in a particular patient and many other features that affect the risk of AF. Given the high-quality indicators of the constructed models, these models may be applicable for early prediction of atrial fibrillation in a patient. The development will allow medical staff to select patients with a high risk of TAF for immediate follow-up or for early radical treatment for TT - surgical treatment or treatment with radioactive iodine instead of long-term medical treatment. We create a scale for assessing the risk of developing Atrial Fibrillation. The main advantage of the scale is its ease of use and good interpretability. Medical staff can use this scale without special software. All constructed models simplify the process of working with patients suffering from thyrotoxicosis. Further work includes improving the quality of the models used by increasing the data features and the use of neural networks, and analysis of atrial fibrillation pathways in dynamics using graphs.

**Acknowledgments.** This research is financially supported by The Russian Science Foundation, Agreement #19-11-00326